\documentclass[aip,reprint,citeautoscript,floatfix]{revtex4-1}
\usepackage{graphicx}
\usepackage{xcolor}
\usepackage{amsmath}
\usepackage{tikz}
\usetikzlibrary{positioning,calc}
\usepackage{adjustbox}
\usepackage[colorlinks=true,linkcolor=blue,citecolor=blue,hypertexnames=false]{hyperref}
\renewcommand{\selectlanguage}[1]{}
\makeatletter
\def\fps@figure{t}
\makeatother
\begin{document}

\title{Differentiable electronic grand canonical learning for electrochemistry}

\author{Phillip E. Pope}
\email{ppope@flatironinstitute.org}
\affiliation{
    Initiative for Computational Catalysis, Flatiron Institute,
    New York, NY 10010, USA
}

\author{Timothy C. Berkelbach}
\email{tberkelbach@flatironinstitute.org}
\affiliation{
    Initiative for Computational Catalysis, Flatiron Institute,
    New York, NY 10010, USA
}
\affiliation{
    Department of Chemistry, Columbia University,
    New York, NY 10027, USA
}

\author{Venkat Kapil}
\email{v.kapil@ucl.ac.uk}
\affiliation{
    Department of Physics and Astronomy, University College London,
    7-19 Gordon St, London WC1H 0AH, UK
}
\affiliation{
    Thomas Young Centre and London Centre for Nanotechnology,
    9 Gordon St, London WC1H 0AH, UK
}

\begin{abstract}
Electronic grand-canonical density functional theory captures how structure and electronic charge respond to electrode potential and, together with statistical mechanics, how these responses give rise to emergent phenomena.
However, its computational cost limits the system sizes and sampling times accessible to first-principles simulations.
Here, we introduce a differentiable architecture that represents the grand potential and obtains forces and electronic charge as its derivatives, exactly satisfying the mixed-derivative constraints required for thermodynamic consistency, and implement it within the equivariant MACE framework as GC-MACE.
We demonstrate more accurate joint learning across multiple electrochemical interfaces and improved transferability to thicker water films relative to previous models, while enabling constant-potential simulations at substantially larger length and time scales than accessible to \textit{ab initio} molecular dynamics.
With broad training data, the framework offers a route to thermodynamically consistent foundation interatomic potentials for the electronic grand canonical ensemble.
\end{abstract}

\maketitle

\newpage

\section{Introduction}

\noindent Simulating electrode-electrolyte interfaces under operating electrochemical conditions reveals the microscopic processes underlying electrocatalysis and battery operation~\cite{tafel_uber_1905, gouy_sur_1910, chapman_li_1913, butler_studies_1924, stern_zur_1924, whittingham_electrical_1976}.
Experiments show that solvation~\cite{toney_voltage-dependent_1994}, pH~\cite{damjanovic_electrode_1967}, and electrode polarization~\cite{vayenas_dependence_1990} strongly affect structure and reactivity at the interface, motivating simulations that capture these effects while allowing exchange of electrons at a prescribed electrode potential.
This electrical boundary condition is naturally described within the electronic grand canonical ensemble~\cite{mermin_thermal_1965}, e.g., using grand-canonical density functional theory (GC-DFT) with a continuum electrolyte~\cite{letchworth-weaver_joint_2012, sundararaman_grand_2017, hormann_grand_2019} or a solvated-jellium treatment of the compensating charge~\cite{kastlunger2018controlled}.
However, the computational cost of GC-DFT limits accessible system sizes and sampling times~\cite{sundararaman_improving_2022, todorova_first-principles_2026, reuter_embrace_2026}, whereas solvent and ion reorganization~\cite{noauthor_structure_1963}, interfacial reactions~\cite{markovic_surface_2002}, and surface reconstruction~\cite{kolb_reconstruction_1996} require large systems and extensive sampling. \\

\noindent Machine learning interatomic potentials (MLIPs) have extended conventional DFT to substantially larger length and time scales by learning potential energy surfaces from first-principles data~\cite{jacobs_practical_2025}.
Modern MLIPs increasingly build physical structure directly into the model, enforcing symmetries such as rotational equivariance and deriving forces from a scalar energy so that they are conservative, while retaining scalability to large datasets and systems~\cite{batatia_design_2022}.
Extending MLIPs to the electronic grand canonical ensemble requires describing the coupled dependence of energy and electronic charge on atomic configuration and electrode potential. \\
\noindent In the electronic grand canonical ensemble (at zero electronic temperature), the thermodynamic potential is the grand potential,
\begin{equation}
    \Omega = E - \mu N_e = E + eUN_e
\end{equation}
where $E$ is the electronic potential energy, $N_e$ is the excess number of electrons compared to an unbiased system, and $\mu$ is the chemical potential, which is fixed by the electrode potential $U$, i.e., $\mu = -eU$ and $e$ is the elementary charge.
Several previous works have developed MLIPs to study systems described by the electrode potential-dependent grand potential.
For example, Ref.~\citenum{tian2025electrochemical} uses separate MLIPs at selected electrode potentials, enabling nanosecond simulations of explicitly solvated interfaces but without a continuous potential dependence within a single model.
Ref.~\citenum{chen2023atomistic} predicts atomic charges and electronegativities through a dual-learning scheme and combines them to construct the grand potential.
Ref.~\citenum{wang2025} uses the electron number $N_e$ as an input and predicts the Fermi level, with a potentiostat adjusting the electron number to regulate the electrode potential.
Finally, Refs.~\citenum{zhou2025,zhou2026} use the electrode potential as an explicit input to predict potential-dependent energies and derive atomic forces, making their potential-conditioning strategy most similar to that adopted here.\\

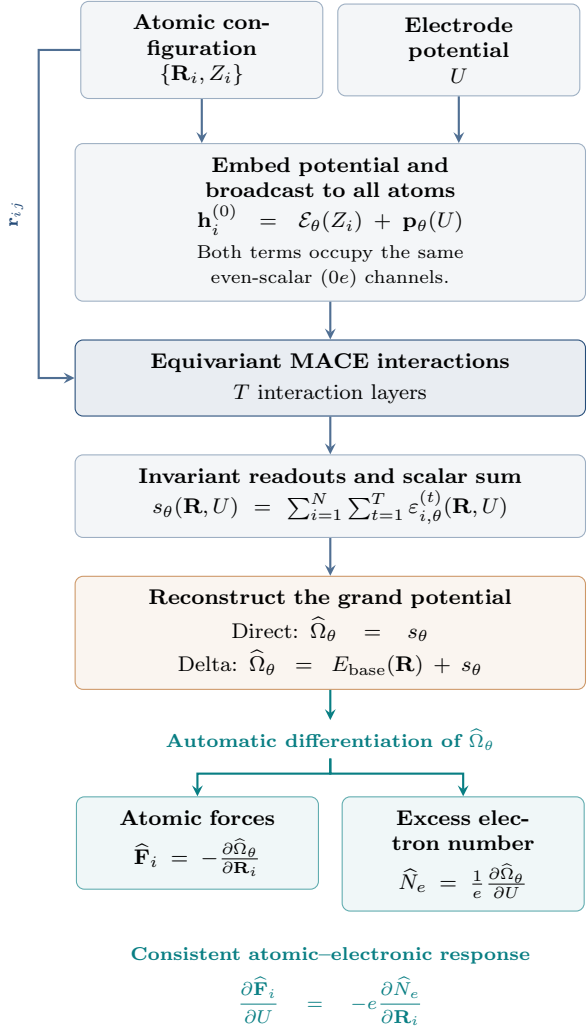
\begin{figure}[!t]
    \centering
    \definecolor{gcblue}{HTML}{244B78}
    \definecolor{gcteal}{HTML}{087E83}
    \definecolor{gcorange}{HTML}{B86B24}
    \begin{adjustbox}{max width=\columnwidth}
    \begin{tikzpicture}[
        font=\footnotesize,
        box/.style={draw=gcblue!45,fill=gcblue!5,rounded corners=3pt,
                    align=center,minimum height=0.78cm,inner sep=5pt},
        wide/.style={box,text width=6.4cm},
        flow/.style={->,>=stealth,thick,draw=gcblue!80},
        response/.style={box,draw=gcteal!65,fill=gcteal!6,
                         text width=2.85cm,minimum height=1cm}
    ]
    \node[box,text width=2.85cm] (atoms) at (-1.7,0)
        {\textbf{Atomic configuration}\\$\{\mathbf R_i,Z_i\}$};
    \node[box,text width=2.85cm] (potential) at (1.7,0)
        {\textbf{Electrode potential}\\$U$};
    \node[wide,below=6mm of $(atoms.south)!0.5!(potential.south)$] (features)
        {\textbf{Embed potential and broadcast to all atoms}\\[2pt]
         $\mathbf h_i^{(0)}=\mathcal E_\theta(Z_i)+\mathbf p_\theta(U)$\\[2pt]
         {\scriptsize Both terms occupy the same even-scalar ($0e$) channels.}};
    \draw[flow] (atoms.south) -- (atoms.south |- features.north);
    \draw[flow] (potential.south) -- (potential.south |- features.north);
    \node[wide,draw=gcblue,fill=gcblue!10,below=5mm of features] (mace)
        {\textbf{Equivariant MACE interactions}\\[2pt]
         $T$ interaction layers};
    \draw[flow] (features) -- (mace);
    \coordinate (geometrytop) at ($(atoms.west)+(-0.55,0)$);
    \coordinate (geometrybottom) at (geometrytop |- mace.west);
    \draw[flow] (atoms.west) -- (geometrytop) -- (geometrybottom) -- (mace.west);
    \node[font=\scriptsize,text=gcblue,rotate=90,anchor=south,inner sep=0pt]
        at ($(geometrytop)!0.5!(geometrybottom)+(-0.20,0)$)
        {$\mathbf r_{ij}$};
    \node[wide,below=5mm of mace] (readout)
        {\textbf{Invariant readouts and scalar sum}\\[2pt]
         $s_\theta(\mathbf R,U)=\sum_{i=1}^{N}\sum_{t=1}^{T}
         \varepsilon_{i,\theta}^{(t)}(\mathbf R,U)$};
    \draw[flow] (mace) -- (readout);
    \node[wide,draw=gcorange!70,fill=gcorange!7,
          below=5mm of readout] (omega)
        {\textbf{Reconstruct the grand potential}\\[3pt]
         Direct: $\widehat\Omega_\theta=s_\theta$\\[3pt]
         Delta: $\widehat\Omega_\theta=E_{\mathrm{base}}(\mathbf R)+s_\theta$};
    \draw[flow] (readout) -- (omega);
    \node[font=\scriptsize,text=gcteal,below=4mm of omega] (autograd)
        {\textbf{Automatic differentiation of $\widehat\Omega_\theta$}};
    \node[response,below=5mm of autograd.south,anchor=north east,xshift=-1.5mm] (forces)
        {\textbf{Atomic forces}\\[4pt]
         $\widehat{\mathbf F}_i=-\frac{\partial\widehat\Omega_\theta}{\partial\mathbf R_i}$};
    \node[response,below=5mm of autograd.south,anchor=north west,xshift=1.5mm] (charge)
       {\textbf{Excess electron number}\\[4pt]
         $\widehat N_e=\frac{1}{e}\frac{\partial\widehat\Omega_\theta}{\partial U}$};
    \draw[flow,draw=gcteal] (omega.south) -- (autograd.north);
    \draw[flow,draw=gcteal] (autograd.south) -- ++(0,-0.2) -| (forces.north);
    \draw[flow,draw=gcteal] (autograd.south) -- ++(0,-0.2) -| (charge.north);
    \node[text width=6.4cm,align=center,text=gcteal,font=\scriptsize,
          below=5mm of $(forces.south)!0.5!(charge.south)$] (identity)
        {\textbf{Consistent atomic--electronic response}\\[4pt]
         $\displaystyle\frac{\partial\widehat{\mathbf F}_i}{\partial U}
         =-e\frac{\partial\widehat N_e}{\partial\mathbf R_i}$};
    \end{tikzpicture}
    \end{adjustbox}
    \caption{
GC-MACE implementation of the differentiable grand-potential architecture.
The species embedding $\mathcal E_\theta(Z_i)$, where $Z_i$ denotes the atomic number, and the final potential embedding $\mathbf p_\theta(U)$ have the same number of even-scalar ($0e$) channels.
The same potential embedding is broadcast to every atom in the configuration and added before the first interaction layer.
Interatomic displacement vectors $\mathbf r_{ij}$ include the appropriate periodic images.
Geometry-dependent equivariant interactions produce invariant atomic contributions $\varepsilon_{i,\theta}^{(t)}$, summed over atoms $i$ and the $T$ interaction layers to obtain the learned scalar $s_\theta$.
This output represents either the total grand potential or its residual relative to a fixed, potential-independent baseline $E_{\mathrm{base}}(\mathbf R)$.
Forces and the excess electron number follow by differentiating the reconstructed grand potential, including the baseline forces in delta learning.
The mixed-derivative relation shown below the outputs holds by construction.
    }
    \label{fig:schematic}
\end{figure}

\noindent We introduce a differentiable architecture implemented with the equivariant MACE framework~\cite{batatia2022mace}, GC-MACE, that learns a single scalar grand potential $\Omega(\mathbf{R},U)$ as a continuous function of atomic positions $\mathbf{R}$ and electrode potential $U$.
Forces and excess electron numbers are obtained as $\mathbf{F}_i=-\partial\Omega/\partial\mathbf{R}_i$ and $N_e=e^{-1}\partial\Omega/\partial U$.
Our architecture improves upon earlier models for the electronic grand canonical ensemble by deriving forces and excess electron numbers from a single grand potential, preserving conservative forces while exactly satisfying the mixed-derivative relationship,
\begin{equation}
\label{eq:gc_mixed_response}
    \frac{\partial \mathbf{F}_i}{\partial U} = -e\frac{\partial N_e}{\partial \mathbf{R}_i},
\end{equation}
which is required for thermodynamic consistency~\cite{mermin_thermal_1965}.
We validate the framework by training a single model across Au, Ag, and Cu interfaces and evaluating its predictions of grand potentials, forces, and excess electron numbers, together with transferability across water-layer thicknesses.
Finally, we revisit the acidified water on Au(111) system, previously studied in \textit{ab initio} studies, by training a single model transferable across five electrode potentials to achieve converged sampling of water layering and potential-induced reorientation in a system containing substantially more explicit water and over much longer timescales.

\section{Results}

\subsection{A differentiable framework for constant-potential learning}
\label{sec:gc_architecture}

\noindent The differentiable architecture takes atomic positions $\mathbf R$, atomic species, and electrode potential $U$ as inputs; here, the potential is referenced to the electrostatic potential deep in the solvent.
In our GC-MACE implementation (Fig.~\ref{fig:schematic}), a continuous embedding of $U$ is projected to match the initial scalar atomic features and added before the equivariant MACE interactions, with scalar atomic readouts summed to obtain the learned grand potential $\widehat\Omega_\theta(\mathbf R,U)$.
Here, $\theta$ denotes the learned parameters, and hats distinguish model predictions from the corresponding unhatted physical quantities.
Conditioning scalar atomic features on the spatial scalar $U$ preserves the spatial transformation properties of the underlying representation while allowing the learned interactions to depend on the electrode potential.
The predicted forces and excess electron numbers follow by automatic differentiation,
\begin{equation}
    \widehat{\mathbf F}_i
    =-\frac{\partial\widehat\Omega_\theta}{\partial\mathbf R_i},
    \qquad
    \widehat N_e
    =\frac{1}{e}
      \frac{\partial\widehat\Omega_\theta}{\partial U}.
    \label{eq:gc_outputs}
\end{equation}
Grand potentials, forces, and excess electron numbers jointly constrain the model through the loss
\begin{equation}
\begin{aligned}
    \mathcal L
    ={}&
    w_\Omega
    \left(
        \frac{\widehat\Omega_\theta-\Omega}{N}
    \right)^2
    +
    \frac{w_F}{3N}
    \sum_{i=1}^{N}
    \left\|
        \widehat{\mathbf F}_i-\mathbf F_i
    \right\|^2
    \\
    &+
    w_{N_e}
    \left(
        \frac{\widehat N_e-N_e}{N}
    \right)^2,
    \label{eq:gc_loss}
\end{aligned}
\end{equation}
where $N$ is the number of atoms, the unhatted targets are supplied by GC-DFT, and $w_\Omega$, $w_F$, and $w_{N_e}$ weight the three contributions.
All three quantities are jointly fit through a common thermodynamic surface. \\

\begin{figure*}[!t]
    \centering
    \includegraphics[width=\textwidth]{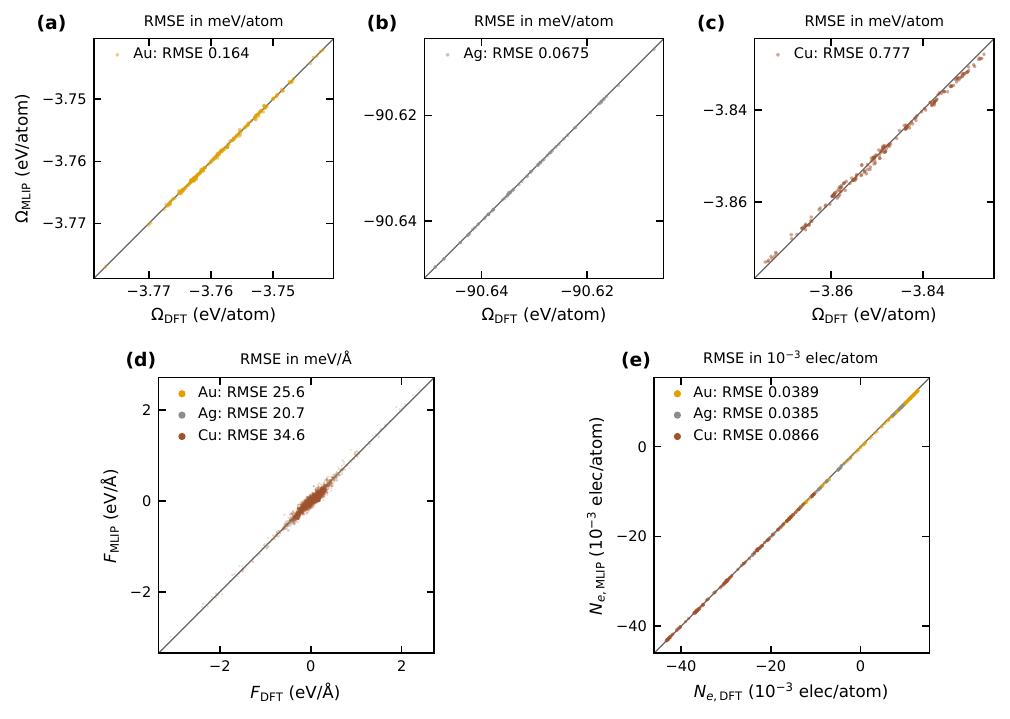}
    \caption{
    Parity between grand-canonical density functional theory (DFT) reference values and predictions from a single GC-MACE machine learning interatomic potential (MLIP), trained directly on total grand potentials using the dataset of Chen \textit{et al.}~\cite{chen2023atomistic}.
    The test set contains $233$, $101$, and $195$ configurations of Au(111), Ag(111), and Cu(111) interfaces, respectively.
    (a--c)~Grand potential per atom, shown separately for Au, Ag, and Cu systems, owing to their different energy scales.
    (d)~Force components and (e)~excess electrons per atom for all slabs.
    The diagonal lines indicate exact agreement.
    Root mean square errors (RMSEs) are reported separately for each metal system in the legends.
    The figure shows predictions from one training run; the main text reports averages over four runs.
    }
    \label{fig:parity-peterson}
\end{figure*}

\noindent For a twice differentiable learned grand potential $\widehat\Omega_\theta$, equality of mixed derivatives guarantees that Eq.~(\ref{eq:gc_mixed_response}) is satisfied.
The electronic response to atomic displacements is therefore consistent with the force response to changes in electrode potential.
Deriving forces from the learned grand potential makes them conservative at fixed potential, while the MACE representation used in GC-MACE preserves translation, rotation, and permutation symmetries.
These properties enable thermodynamically consistent constant-potential dynamics without separately predicting or adjusting the electron number.\\

\noindent We train the model either directly on GC-DFT grand potentials or on their residuals relative to a pretrained model at the potential of zero charge.
The residual formulation preserves the same derivative relations.

\subsection{Au, Ag, and Cu surface chemistry}
\label{sec:benchmark}

\begin{figure*}[!t]
    \centering
    \includegraphics[width=\textwidth]{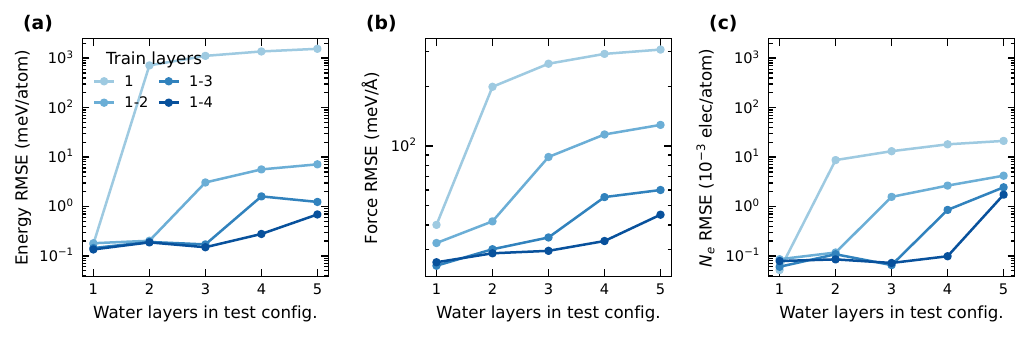}
    \caption{
    Extrapolation across water layers using the dataset of Chen \textit{et al.}~\cite{chen2023atomistic}, with GC-MACE trained on total grand potentials.
    Models trained on configurations with up to 1--4 water layers (line shading) are evaluated on configurations with 1--5 layers: (a)~grand potential, (b)~force, and (c)~excess-electron root mean square errors (RMSEs) versus the number of water layers in the test configurations.
    }
    \label{fig:extrapolation}
\end{figure*}

\noindent We first assess the differentiable architecture through its GC-MACE implementation on the dataset of Chen et al.~\cite{chen2023atomistic}, which comprises two subsets with grand potentials, forces, and excess electron numbers computed using GC-DFT with the solvated jellium method (SJM)~\cite{kastlunger2018controlled} implemented in GPAW~\cite{mortensen2024gpaw}.
The first subset contains $5293$ configurations of Au(111), Ag(111), and Cu(111) electrochemical interfaces, obtained from nudged elastic band calculations of (1) the Volmer and Heyrovsky steps of the hydrogen evolution reaction on Au and Ag surfaces at electrode potentials spanning $4.0$--$4.85$~V and (2) CO chemisorption on Cu at electrode potentials spanning $3.79$--$4.61$~V.
Over this potential range, the excess electron number per atom ranges from approximately $-40\times10^{-3}$ to $20\times10^{-3}$.
We divide these configurations into training, validation, and test sets containing $4235$, $529$, and $529$ configurations, respectively.
The second subset contains $3015$ Au(111) configurations with water films ranging from one to five layers and four water molecules per layer, with $250$--$896$ configurations at each thickness.
For the surface-chemistry benchmark, we train a single GC-MACE model directly on the combined Au, Ag, and Cu training configurations. \\

\noindent Figure~\ref{fig:parity-peterson} compares GC-MACE predictions of grand potentials, forces, and excess electron numbers with GC-DFT reference values on the test set.
For Au, Ag, and Cu, respectively, we obtain grand-potential root mean square errors (RMSEs) of $0.161\pm0.003$, $0.069\pm0.003$, and $0.781\pm0.002$~meV/atom, force RMSEs of $25.7\pm0.4$, $20.5\pm0.3$, and $36.4\pm0.7$~meV/\AA, and excess-electron RMSEs of $(0.038,\,0.037,\,0.088)\times10^{-3}$~$e$/atom, each with a standard error of $0.001\times10^{-3}$~$e$/atom.
Quoted uncertainties are standard errors of the mean over four training runs differing only in random seed; Fig.~\ref{fig:parity-peterson} shows one run.
By comparison, Ref.~\citenum{chen2023atomistic} trains separate models for each metal and learns grand potential and excess charge through distinct regressions, without force training in this benchmark.
For Au(111), Ag(111), and Cu(111), those models yield grand-potential RMSEs of $0.15$, $0.21$, and $0.24$~meV/atom and excess-electron RMSEs of $(0.058,\,0.050,\,0.101)\times10^{-3}$~$e$/atom, respectively~\cite{chen2023atomistic}.
Comparing the two models, we find both models achieve sub-meV energy errors but neither outperforms the other across all elements: GC-MACE is $3\times$ more accurate on Ag, $7\%$ less accurate on Au, and $3\times$ less accurate on Cu.
GC-MACE has lower reported excess-electron RMSEs on all three metals and simultaneously fits forces, which were not included in training the reference models for this benchmark.
It also guarantees the derivative relations between atomic and electronic responses. \\
\noindent Figure~\ref{fig:extrapolation} assesses transferability with solvent thickness by training GC-MACE on interfaces containing up to one, two, three, or four water layers and evaluating on systems containing as many as five layers, following Ref.~\citenum{chen2023atomistic}.
Both approaches obtain large errors when training on one layer and testing on the five-layer system.
GC-MACE has grand-potential, force, and excess-electron RMSEs of $1551$~meV/atom, $305$~meV/\AA{}, and $21\times10^{-3}$~$e$/atom, respectively.
Including two layers in training reduces the GC-MACE grand-potential and excess-electron RMSEs to $7.1$~meV/atom and $4.2\times10^{-3}$~$e$/atom, respectively, whereas Ref.~\citenum{chen2023atomistic} reports energy errors of order $100$--$1000$~meV/atom and charge errors of order $10^{-3}$--$10^{-2}$~$e$/atom for thicker interfaces.
With three- and four-layer training, the GC-MACE five-layer errors decrease further to $1.2$ and $0.69$~meV/atom in the grand potential and to $2.4$ and $1.7\times10^{-3}$~$e$/atom in excess electron number.
For test configurations containing one to four water layers, the model trained on configurations with up to four layers has grand-potential RMSEs of $0.135$--$0.278$~meV/atom.
By comparison, the model of Ref.~\citenum{chen2023atomistic} reaches its low-error five-layer regime only when films of comparable thickness are represented in training.
These results indicate that GC-MACE exhibits improved transferability to thicker solvent films, most clearly in the grand potential, while also evaluating force transferability with forces and electronic response derived from the same thermodynamic surface.

\subsection{Acidified water on Au(111)}
\label{sec:water}

\begin{figure*}[!t]
    \centering
    \includegraphics[width=\textwidth]{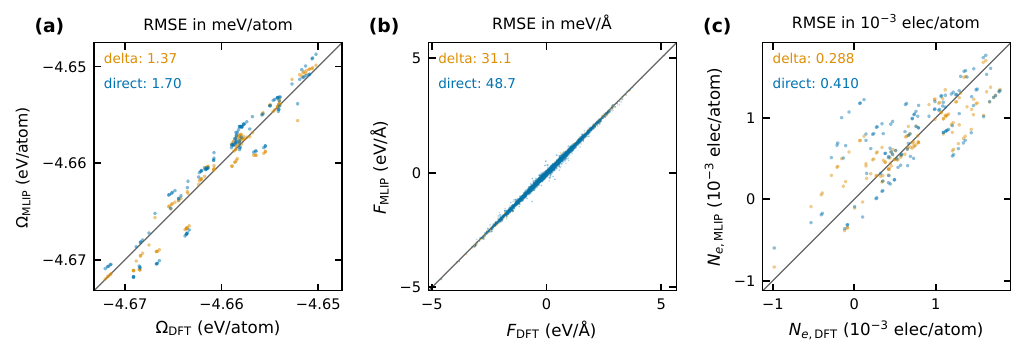}
    \caption{
    Parity for the differentiable architecture on our grand-canonical reference dataset from density functional theory (DFT) calculations for acidified water on Au(111) (six water layers), evaluated on the held-out test split ($125$ configurations, $3.5$--$5.5$~V).
    Predictions from machine learning interatomic potentials (MLIPs) trained on delta residuals (orange) and total grand potentials (blue) are compared on shared axes showing total quantities.
    Molecular dynamics uses the delta model.
    The columns show the grand potential per atom, the force components, and the excess electrons per atom; root mean square errors (RMSEs) are annotated in each panel.
    Delta predictions are converted to total grand potentials and forces by adding back the same per-frame baseline used to construct the residual targets, so the total-quantity errors equal the residual errors.
    }
    \label{fig:parity-md}
\end{figure*}

\noindent We now turn to our main system of interest, acidified water on Au(111).
As discussed in more detail in Sec.~\ref{sec:gcdft}, we generated a GC-DFT dataset of $127$-atom configurations comprising a three-layer $3\times2$ Au(111) slab ($18$ atoms), $36$ water molecules arranged in six layers, and one excess proton.
The electrode potentials span $3.5$--$5.5$~V.
The training, validation, and test sets contain $2870$, $163$, and $125$ configurations, respectively.
For this system, we compare two models: one that directly learns the total grand potential and one that uses delta learning relative to a potential-independent foundation-model baseline (Sec.~\ref{sec:arch}); we will refer to these as the ``direct'' and ``delta'' learned models, respectively. \\

\begin{figure*}[!t]
    \centering
    \includegraphics[width=\textwidth]{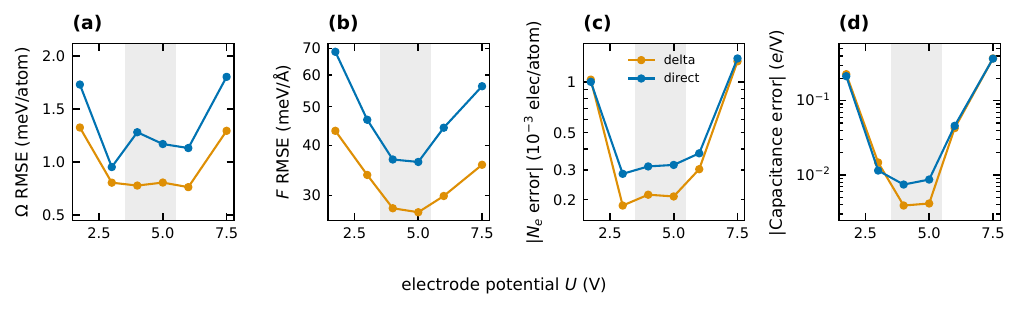}
    \caption{
    Extrapolation in the electrode potential.
    Both models were trained only on configurations with $3.5<U<5.5$~V (shaded) and are evaluated against grand-canonical density functional theory reference values for $281$ further configurations of the same system spanning $1.2$--$8.4$~V, none of which appear in the training, validation, or test splits ($83$ below the window, $102$ inside, $96$ above).
    Every panel reports an error, averaged within potential bins: (a)~grand-potential root mean square error (RMSE); (b)~force RMSE; (c)~mean absolute error in the excess electron number per atom; and (d)~mean absolute error in the differential capacitance $C=-e\,\mathrm{d}N_e/\mathrm{d}U$.
    The capacitance is taken \emph{within} a configuration, from $122$ pairs that share a geometry and differ only in potential ($37$ geometries, gaps $0.1$--$2.0$~V), so that model and reference are differenced over identical pairs and the comparison is free of geometry-to-geometry variation.
    Note the logarithmic scales on (c) and (d): the four quantities leave the training window at very different rates.
    }
    \label{fig:potential-extrapolation}
\end{figure*}

\noindent Figure~\ref{fig:parity-md} compares GC-MACE predictions with GC-DFT reference values on the test set.
The delta model gives grand-potential, force, and excess-electron RMSEs of $1.37$~meV/atom, $31.1$~meV/\AA{}, and $0.29\times10^{-3}$~$e$/atom, respectively, compared with the direct model, which gives RMSEs of $1.70$~meV/atom, $48.7$~meV/\AA{}, and $0.41\times10^{-3}$~$e$/atom.
Thus the delta model outperforms its direct counterpart on all three quantities. \\

\noindent Figure~\ref{fig:potential-extrapolation} tests extrapolation in electrode potential by comparing the models trained over $3.5$--$5.5$~V with GC-DFT reference data spanning $1.2$--$8.4$~V.
For the delta model, predicting outside the training window raises the grand-potential RMSE from $0.79$~meV/atom inside to $1.20$ below the window and $1.12$ above.
Likewise the force RMSE increases from $27.6$~meV/\AA{} inside the window to $40.9$~meV/\AA{} below and $33.7$~meV/\AA{} above.
The excess electron predictions degrade by a factor of four to five, from $0.29\times10^{-3}$~$e$/atom inside the window to
$1.25\times10^{-3}$ below and $1.45\times10^{-3}$ above.
The differential capacitance $C=-e\,dN_e/dU$, which should be roughly independent of $U$ near the potential of zero charge, does not extrapolate well to potentials outside of the training data.
Inside the window, the model reproduces $C$ to $0.005$~$e$/V against a reference value of $0.041$~$e$/V, but outside the window the reference rises roughly fivefold, to $0.227$~$e$/V below and $0.215$~$e$/V above, while the model captures almost none of the increase, resulting in errors of $0.20$ and $0.19$~$e$/V, approximately the same magnitude as the reference signal.
The same ordering holds for the direct model trained on total grand potentials.
Thus the degradation reflects the limits of the training window and the nonlinear electronic response at large applied potentials rather than the choice of learning target.
These results show that the electronic response, and especially its derivative, is far more sensitive to potential extrapolation than the grand potential or the forces, so small grand-potential errors alone do not establish the accuracy of the predicted charge response. \\

\noindent Having evaluated the generalization performance of our model, we now assess its molecular dynamics (MD) capabilities, a more demanding test with greater practical relevance.
A similar system was previously studied using potential-dependent \textit{ab initio} MD (AIMD) simulations~\cite{gunasekaran2024water}, although our study differs in interfacial geometry, electrolyte composition, and electrochemical boundary conditions.
Our SJM reference calculations impose the electrode potential at fixed ionic composition, allowing the electron number to adjust, with the countercharge represented by a jellium region screened by a dielectric continuum~\cite{kastlunger2018controlled}. \\
\noindent Our 508-atom system represents a single half-cell comprising a three-layer $6\times4$ Au(111) slab, 144 water molecules arranged in six layers, and four excess protons.
We run trajectories at $330$~K at five potentials ranging from $-0.59$ to $0.95$~V vs the standard hydrogen electrode (SHE) using the same GC-MACE model throughout (we reference our absolute electrode potentials to the SHE by subtracting $4.44$~V).
Each trajectory provides $350$~ps of sampling, a duration significantly longer than that accessible in standard AIMD~\cite{gunasekaran2024water}.
The simulations use \textsc{torch-sim}~\cite{cohen2025torchsim} with a BAOAB Langevin integrator at a friction of $10^{-2}$ (a thermostat time constant of $1.018$~ps) and a $0.25$~fs timestep, with further MD settings given in Sec.~\ref{sec:methods}.\\

\begin{figure*}
    \centering
    \includegraphics[width=\textwidth]{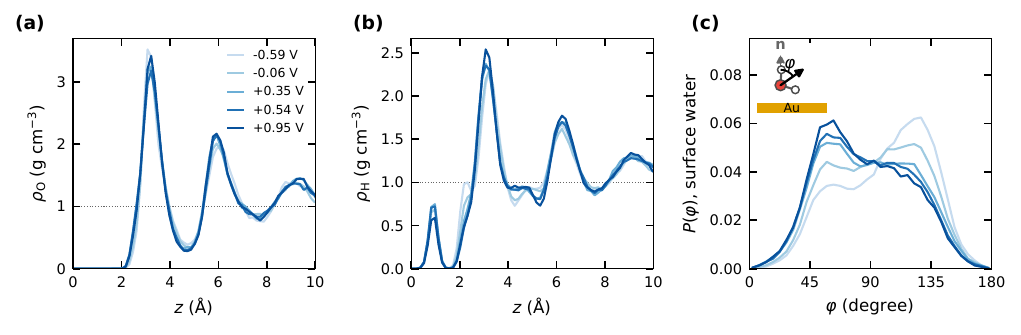}
    \caption{
    Structure of interfacial water at constant electrode potential: GC-MACE molecular dynamics
    ($350$~ps at each of five electrode potentials, $330$~K, delta model with D3 dispersion, each trajectory started by relaxing the same initial structure at its own potential):
    Left and middle columns: density profiles calculated from oxygen and hydrogen positions, respectively, versus height $z$ above the outermost Au layer, assigning one water-molecule mass to each oxygen and half a water-molecule mass to each hydrogen.
    Right column: the orientation distribution of the water bisector relative to the surface normal for surface water ($z<4$~\AA), showing the potential-induced reorientation.
    The inset defines $\varphi$ as the angle between the outward surface normal $\mathbf{n}$ and the water bisector, which points from the oxygen towards the two hydrogens (i.e.\ along the molecular dipole): $\varphi=0$ has both O--H bonds directed away from the electrode.
    }
    \label{fig:density}
\end{figure*}

\noindent Figure~\ref{fig:density} compares the oxygen and hydrogen density profiles and orientation distributions of interfacial water.
Water-density profiles and orientation distributions obtained using the potential-independent \texttt{mace-mh-1.model} baseline with D3(BJ) dispersion are shown in Sec.~\ref{app:mh1-water} (Fig.~\ref{fig:si-density-mh1}) of the Supplementary Information.
We observe pronounced water layering near the electrode, but with a relatively weak potential dependence.
The hydrogen-density peak closest to the Au surface is slightly reduced at the most positive potentials.
In contrast, we observe a relatively strong dependence of the water orientation at the surface: the mean angle between the water bisector and the surface normal decreases monotonically with increasing potential, consistent with an increased attraction between the negatively charged oxygen and the positively charged surface.
The same qualitative behavior was observed in Ref.~\citenum{gunasekaran2024water}.
The potentials shown are referenced to the standard hydrogen electrode.

\section{Conclusions}

\noindent We have introduced a differentiable architecture that learns the grand potential of electrochemical interfaces as a continuous function of atomic configuration and electrode potential, obtaining forces and excess electron numbers by differentiation, and demonstrated it through the equivariant GC-MACE implementation.
This construction ensures conservative forces and exactly satisfies the mixed-derivative relations between force responses to electrode potential and charge responses to atomic displacements, establishing thermodynamic consistency between atomic and electronic responses.
Differential capacitance follows from the curvature of the same grand-potential surface. \\

\noindent When benchmarked on the GC-DFT data of Ref.~\citenum{chen2023atomistic}, a single GC-MACE model accurately reproduces grand potentials, forces, and excess electron numbers across electrochemical Au, Ag, and Cu interfaces.
These results demonstrate that the differentiable formulation can describe multiple electrode materials within a common, thermodynamically consistent model. \\

\noindent For acidified water on Au(111), we showed that model predictions worsen at electrode potentials outside the training range, possibly because of breakdowns in the linear response relationship between excess charge and electrode potential.
We used a single GC-MACE model to perform $350$~ps trajectories at each of five electrode potentials in a 508-atom half-cell.
The simulations resolve interfacial water layering and potential-induced reorientation over sampling times substantially longer than those of a previous AIMD study~\cite{gunasekaran2024water}.
Future work will explore the performance of our model with an explicit electrolyte, whose precise behavior may not be captured by the dielectric continuum treatment used here. \\
\noindent By deriving atomic and electronic responses from a single learned grand-potential surface, the differentiable formulation enforces consistency between them.
In its current GC-MACE implementation, this construction is combined with equivariant atomic representations and multi-element modelling.
The formulation could also be incorporated into other differentiable architectures used to build foundation MLIPs~\cite{jacobs_practical_2025, yuan_foundation_2026, creed_six_2026}.
With sufficiently broad GC-DFT training sets for solvated and electrified catalytic interfaces, analogous in scope to the Open Catalyst 2020~\cite{chanussot_open_2021}, Open Catalyst 2022~\cite{tran_open_2023}, and Open Catalyst 2025~\cite{sahoo_open_2025} datasets for conventional DFT, this approach could support foundation MLIPs transferable across electrode potentials, electrodes, adsorbates, and electrolytes.

\section{Methods}
\label{sec:methods}

\subsection{GC-MACE implementation details}
\label{sec:gc_implementation}
\noindent For the GC-MACE implementation presented here, we use equivariant MACE interactions~\cite{batatia2022mace}, built on the Atomic Cluster Expansion~\cite{drautz2019atomic}.
Interatomic displacements within a $6.0$~\AA{} cutoff are expanded using eight Bessel radial functions with a polynomial cutoff of order $p=5$ and spherical harmonics up to $\ell=3$.
The model uses two residual interaction layers with body order four (correlation order three) and hidden features $128\times0e+128\times1o+128\times2e$ ($128$ channels, $L_{\max}=2$).
The radial multilayer perceptron has three hidden layers of $64$ units each.
The electrode potential is embedded using two linear layers of width $1024$, separated by a sigmoid linear unit (SiLU) activation, followed by a linear projection to $128$ channels and a second SiLU activation.
The resulting embedding is shared by all atoms in a configuration and added to their initial $128$-channel scalar species features.
Energy predictions are obtained using a linear readout at  intermediate layers and a gated multilayer perceptron with $16\times0e$ hidden features at the final layer.
Atomic reference energies are averages over the training set, and the model output scaling is set using the root mean square training forces.
These settings yield a model with approximately $2.35\times10^6$ parameters.%

\subsection{GC-MACE training for direct and delta learning}
\label{sec:arch}

\noindent Using the architecture described in Sec.~\ref{sec:gc_implementation}, we optimize the joint loss in Eq.~\eqref{eq:gc_loss} using Adam with AMSGrad, $\beta_1=0.9$, weight decay $5\times10^{-7}$, and gradient clipping at $10$.
The architecture and optimization settings are shared by every model reported here, with the Chen \textit{et al.} benchmark using the different batch size and loss weights specified below.
Training proceeds for $500$ epochs, with a second stage beginning at epoch $250$.
The initial learning rate is $5\times10^{-3}$ and is reduced on plateau by a factor of $0.8$ with a patience of $50$ epochs; it is set to $10^{-3}$ at the start of the second stage.
For acidified water, the loss weights $(w_\Omega,w_F,w_{N_e})$ are $(1,10,1.613\times10^5)$ in the first stage and $(1,100,1.613\times10^6)$ in the second, with a batch size of $11$.
At the start of the second stage, the force and excess-electron weights are raised tenfold relative to the grand-potential weight, which is held fixed.
For the Chen \textit{et al.} benchmark, the weights remain $(1600,1,4\times10^4)$ throughout training, with a batch size of $16$.
These weights are chosen so that each of the three terms contributes comparably to the total loss.
Training uses float64 precision, and reported predictions use exponentially averaged model weights with a decay of $0.99$. \\

\noindent In direct learning, the learned scalar represents the total grand potential $\widehat\Omega_\theta(\mathbf R,U)$.
Its derivatives give the predicted forces and excess electron numbers through Eq.~\eqref{eq:gc_outputs}, and all three quantities are jointly fit to their GC-DFT reference values using Eq.~\eqref{eq:gc_loss}.
This formulation is used for the Chen \textit{et al.} benchmark in Sec.~\ref{sec:benchmark} and for acidified water in Sec.~\ref{sec:water}. \\

\noindent In delta learning, the learned scalar instead represents the grand-potential residual relative to a fixed, potential-independent baseline,
\begin{equation}
    \Delta\Omega(\mathbf R,U)
    =
    \Omega(\mathbf R,U)-E_{\mathrm{base}}(\mathbf R).
    \label{eq:gc_delta_target}
\end{equation}
The same derivative construction and joint loss are used, with the grand-potential and force targets replaced by $\Delta\Omega$ and $\Delta\mathbf F_i=\mathbf F_i-\mathbf F_{i,\mathrm{base}}$, where $\mathbf F_{i,\mathrm{base}}=-\partial E_{\mathrm{base}}/\partial\mathbf R_i$.
The excess-electron targets remain unchanged because the baseline is independent of electrode potential.
The total predictions are reconstructed as
\begin{equation}
\begin{aligned}
    \widehat\Omega_\theta
    &=E_{\mathrm{base}}+\widehat{\Delta\Omega}_\theta,
    \\
    \widehat{\mathbf F}_i
    &=\mathbf F_{i,\mathrm{base}}
      -\frac{\partial\widehat{\Delta\Omega}_\theta}
             {\partial\mathbf R_i},
    \\
    \widehat N_e
    &=\frac{1}{e}
      \frac{\partial\widehat{\Delta\Omega}_\theta}{\partial U}.
    \label{eq:gc_delta_outputs}
\end{aligned}
\end{equation}
This formulation is used for the acidified-water MD presented in Sec.~\ref{sec:water}.
\noindent The baseline used for delta learning is the pretrained mh-1 foundation MLIP (\texttt{mace\_mp}, checkpoint \texttt{mace-mh-1.model}, head \texttt{oc20\_usemppbe}).
The mh-1 baseline is evaluated without dispersion corrections when constructing the residual grand-potential and force targets.
For MD we add the D3 dispersion energy and its corresponding forces to the reconstructed grand potential and forces.

\subsection{Physical symmetries and derivative relations}

\noindent Translation invariance of the learned grand potential is inherited by its first and mixed derivatives.
This guarantees the translation invariance of the predicted charge and two sum rules:
\begin{equation}
 \sum_{i=1}^{N}\widehat{\mathbf F}_i = \mathbf 0,\quad\quad
    \sum_{i=1}^{N} \frac{\partial\widehat N_e}{\partial\mathbf R_i}
    =\mathbf{0}.
\end{equation}
At fixed electrode potential, classical Hamiltonian dynamics on this surface conserves $K+\widehat\Omega_\theta$, where $K$ is the classical nuclear kinetic energy.
\noindent Finally, at fixed atomic positions, the electronic differential capacitance is
\begin{equation}
    C_{\mathrm{el}}
    =
    \left.\frac{\partial Q}{\partial U}\right|_{\mathbf R}
    =
    -e\left.\frac{\partial N_e}{\partial U}\right|_{\mathbf R}
    =
    -\left.
      \frac{\partial^2\Omega}{\partial U^2}
      \right|_{\mathbf R},
    \label{eq:gc_capacitance}
\end{equation}
where $Q=-eN_e$ is the excess electronic charge.
The model prediction follows by replacing $\Omega$ with $\widehat\Omega_\theta$.

\subsection{GC-DFT reference data}
\label{sec:gcdft}

\paragraph{Electronic structure.}

\noindent Dataset compositions, electrode-potential ranges, and training, validation, and test splits are described in Secs.~\ref{sec:benchmark} and \ref{sec:water}.
Reference-data distributions for both datasets are shown in Sec.~\ref{app:data-distributions} (Fig.~\ref{fig:si-data-dist}) of the Supplementary Information.
Our GC-DFT calculations use the PBE functional and the SJM~\cite{kastlunger2018controlled} implemented in GPAW~\cite{mortensen2024gpaw} and driven through ASE~\cite{ase-paper}.
The metal slab, adsorbates, and any explicit solvent molecules are surrounded above by a dielectric continuum, with a faraway jellium slab carrying any necessary compensating charge.
We apply a dipole correction in the surface plane, and the electrode potential is referenced to the electrostatic potential deep in the dielectric continuum. \\
\noindent The following settings are used for every GC-DFT reference label.
The calculations use GPAW in finite-difference mode.
The grid is set from a target spacing of $0.1$~\AA{}, with the number of grid points rounded up to a multiple of $16$ in each direction.
For the acidified-water system, the resulting coarse and fine grids contain $80\times48\times608$ and $160\times96\times1216$ points, respectively.
Brillouin-zone sampling uses a $4\times6\times1$ Monkhorst--Pack mesh.
Fermi--Dirac occupations have a width of $0.1$~eV, and the force convergence tolerance is $10^{-3}$~eV/\AA{}.
The implicit solvent uses the smooth-cavity model of Held and Walter~\cite{held2014simplified}.
The cavity uses a Power12 effective potential with $u_0=0.18$~eV.
The dielectric is linear, with $\varepsilon_\infty=78.36$, and the surface tension is $18.4$~mN/m.
The cavity construction includes ghost atoms below each oxygen atom at a spacing of $2/3$~\AA{}.
The compensating jellium region extends from above the water film to the cell boundary. \\
\paragraph{Dataset sampling}

\noindent Configurations for the initial acidified-water dataset were selected from MD driven by the potential-independent foundation MLIP.
These initial configurations were drawn at random from the trajectory rather than at fixed intervals, and no two selected configurations are structurally close.
Target electrode potentials in the range $3.5$--$5.5$~V were randomly assigned to each configuration after selection.
The electrode potentials used in the dataset are those achieved by the converged SJM calculation, which may differ from the target potential up to a tolerance of $0.01$~V.
The initial GC-DFT dataset comprised $834$ geometries.
However, this dataset required expansion because MD with MLIPs trained on it exhibited short-range O--H instabilities and spontaneous formation of hydronium--hydroxide pairs at a much higher frequency than physically expected.
These problems were addressed through short-range data augmentation and active-learning procedures, as detailed in Sec.~\ref{app:al} of the Supplementary Information.
The final GC-DFT reference dataset contained $904$ geometries and $3158$ labelled rows, an increase of $70$ geometries. \\

\noindent For the dataset of Chen \textit{et al.}~\cite{chen2023atomistic}, we use the same $80/10/10$ training, validation, and test split as the original study, obtained by randomly shuffling configurations using a fixed seed.
The acidified-water dataset was instead split by geometry using a fixed random seed, with all entries sharing the same atomic geometry assigned to the same training, validation, or test set.

\subsection{Constant-potential molecular dynamics and analysis}

\noindent MD uses an in-house implementation built upon \textsc{torch-sim}~\cite{cohen2025torchsim} and \textsc{torch-d3}~\cite{takamoto_towards_2022}, with the reconstructed forces given in Eq.~\eqref{eq:gc_delta_outputs}.
The simulations use the $NVT$ ensemble with a BAOAB Langevin thermostat~\cite{leimkuhler_robust_2013} and a conservative $0.25$~fs timestep.
The friction is $10^{-2}$ in the integrator's internal units, corresponding to $9.8\times10^{-4}$~fs$^{-1}$ and a damping time of $1.018$~ps.
The potentials, trajectory duration, temperature, integration parameters, and equilibration period are given in Sec.~\ref{sec:water}.
The simulation cell measures $17.31\times9.99\times60.71$~\AA{}, with periodic boundaries in the surface plane and open boundaries along the surface normal.
The bottom Au layer, containing $24$ atoms, is fixed.
The initial structure is constructed on a lattice by arranging $36$ water molecules in six layers spaced $4$~\AA{} apart above the Au(111) slab, adding one excess proton, and tiling the cell $2\times2$ in the surface plane, giving $144$ water molecules and four excess protons.
Each trajectory starts from the same initial atomic structure, relaxed at its assigned electrode potential.
Relaxation uses Broyden-Fletcher-Goldfarb-Shanno optimization with a maximum force threshold of $0.01$~eV/\AA{}, followed by Maxwell--Boltzmann initial velocities at $330$~K.
Trajectories of $350$~ps are generated at electrode potentials of $3.85$, $4.38$, $4.79$, $4.98$, and $5.39$~V, using a single thermostat seed per potential.
D3 dispersion uses Becke--Johnson damping with PBE parameters and a cutoff of $40\,a_0=21.17$~\AA{}.
MD is performed in float32 precision, with configurations saved every $0.1$~ps. \\

\noindent Oxygen and hydrogen density profiles are obtained by binning atomic heights relative to the outermost Au layer and averaging over the retained trajectory.
The outermost Au plane is recomputed each frame, and heights are binned at $0.15$~\AA{}.
Mass densities are normalised by the lateral cell area and bin width.
Both profiles are reported as equivalent water mass densities, so that oxygen counts carry a full water mass and hydrogen counts half of one; this matches the convention of the reference AIMD profiles and makes $\rho_\mathrm{O}$ and $\rho_\mathrm{H}$ directly comparable.
Water orientation is measured by the angle between the outward surface normal and the water bisector, directed from oxygen towards the two hydrogen atoms.
Orientation distributions include molecules whose oxygen atoms lie within $4$~\AA{} of the outermost Au layer.
Orientations are histogrammed in $5^\circ$ bins. \\

\noindent Molecules are identified per frame rather than by a fixed assignment.
Each hydrogen is attached to its nearest oxygen under the minimum-image convention in the two periodic directions, and counted as bonded if the separation is below $1.30$~\AA{}; a water molecule is an oxygen carrying exactly two such hydrogens.
An oxygen carrying three hydrogens (hydronium) or one hydrogen (hydroxide) has no well-defined bisector and is omitted from the orientation histogram for that frame alone, while still contributing to the density profiles.

\section*{Acknowledgements}

\noindent We thank Xi Chen and Andrew Peterson for helpful discussions and for sharing the data of Ref.~\citenum{chen2023atomistic}.
We thank Stefano Falletta and William Baldwin for insightful discussions.
The Flatiron Institute is a division of the Simons Foundation.

\section*{Model and data availability}

Models and data will be made available upon publication.

\clearpage
\onecolumngrid
\pagenumbering{arabic}
\setcounter{section}{0}
\setcounter{subsection}{0}
\setcounter{figure}{0}
\setcounter{table}{0}
\setcounter{equation}{0}
\renewcommand{\thesection}{S\arabic{section}}
\renewcommand{\thefigure}{S\arabic{figure}}
\renewcommand{\thetable}{S\arabic{table}}
\renewcommand{\theequation}{S\arabic{equation}}
\renewcommand{\theHsection}{supp.\arabic{section}}
\renewcommand{\theHsubsection}{supp.\arabic{section}.\arabic{subsection}}
\renewcommand{\theHfigure}{supp.\arabic{figure}}
\renewcommand{\theHtable}{supp.\arabic{table}}
\renewcommand{\theHequation}{supp.\arabic{equation}}
\section*{Supplementary Information}
\subsection*{Contents}
\noindent\hyperref[app:data-distributions]{\ref*{app:data-distributions}\quad Reference-data distributions}\dotfill\pageref{app:data-distributions}\\
\hyperref[app:al]{\ref*{app:al}\quad Additional data collection}\dotfill\pageref{app:al}\\
\hspace*{1.5em}\hyperref[si:short-range-augmentation]{\ref*{si:short-range-augmentation}\quad Short-range Data Augmentation}\dotfill\pageref{si:short-range-augmentation}\\
\hspace*{1.5em}\hyperref[si:active-learning]{\ref*{si:active-learning}\quad Active Learning}\dotfill\pageref{si:active-learning}\\
\hyperref[app:mh1-water]{\ref*{app:mh1-water}\quad Water structure using the mace-mh-1 foundation MLIP}\dotfill\pageref{app:mh1-water}

\clearpage
\section{Reference-data distributions}
\label{app:data-distributions}

\begin{figure}[!htbp]
    \centering
    \includegraphics[width=\textwidth]{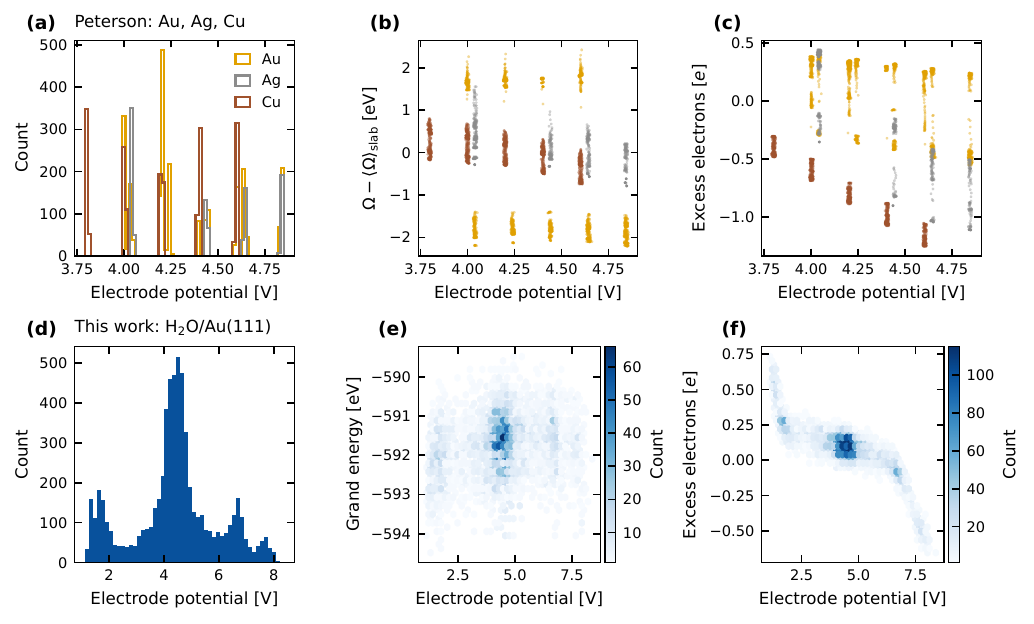}
    \caption{
    Distributions of the grand-canonical density functional theory (GC-DFT) reference data.
    (a--c)~dataset of Chen \textit{et al.}~\cite{chen2023atomistic} (Au, Ag, Cu slabs): (a)~electrode potential, (b)~grand energy versus potential with the per-slab mean removed (the slabs' absolute energy scales differ), and (c)~excess electrons versus potential.
    (d--f)~Our GC-DFT dataset of acidified water on Au(111) (six water layers): (d)~electrode potential, and joint distributions of (e)~grand energy and (f)~excess electrons versus potential.
    }
    \label{fig:si-data-dist}
\end{figure}

Figure~\ref{fig:si-data-dist} shows the distributions of the reference data.

\clearpage
\section{Additional data collection}
\label{app:al}

\noindent In the course of our investigation we found that accurate fitting of the training and test data did not always guarantee stable or physical molecular dynamics (MD).
Earlier iterations of the model reached grand-potential, force and excess-electron errors comparable to those reported in Sec.~II\,C, yet produced several distinct failures in MD.
Two of these were repaired by collecting more data, and are described here: (i) non-physical short-range interactions, in which a short O--H contact led to a divergence in energy; and (ii) spontaneous formation of $\mathrm{H_3O^+}$/$\mathrm{OH^-}$ pairs, which are absent from the training data.
The two were addressed differently: the first by short-range data augmentation, the second by collecting additional GC-DFT labels.
We describe each in turn.

\subsection{Short-range Data Augmentation}
\label{si:short-range-augmentation}

\noindent The short-range failure (i) was addressed by augmenting the dataset with "compressed" versions of existing GC-DFT labels, where the compression energy is calculated using the foundation machine learning interatomic potential (MLIP) \texttt{mace-mh-1.model} (head \texttt{oc20\_usemppbe}) with its potential-independent Foundation Embeddings.
This baseline does not exhibit short-range failures.
Writing $\mathbf{R}$ for the coordinates of a labelled parent frame and $\mathbf{R}'$ for a compressed copy of it, the grand potential assigned to the copy is
\begin{equation}
\Omega_{\mathrm{aug}}(\mathbf{R}'; U) = \Omega_{\mathrm{DFT}}(\mathbf{R}; U)
  + E_{\mathrm{base}}(\mathbf{R}') - E_{\mathrm{base}}(\mathbf{R}),
\label{eq:sraug}
\end{equation}
where $E_{\mathrm{base}}$ is the energy predicted using these potential-independent Foundation Embeddings and $\Omega_{\mathrm{DFT}}$ the grand-canonical reference label.
The electrode potential $U$ is unchanged by the compression, and $E_{\mathrm{base}}$ depends on coordinates alone, so the difference in Eq.~(\ref{eq:sraug}) carries no potential dependence and the augmented label only inherits potential dependence from the parent.
This removed the failure, and all models reported here are trained on the augmented set.

\subsection{Active Learning}
\label{si:active-learning}

\noindent In total $197$ additional GC-DFT labels across $70$ geometries were collected, under the same GC-DFT protocol as the base set.
With the $834$ geometries of the base set these give the $904$ geometries and $3158$ labelled rows of the final dataset.
Frames were selected by criteria fixed in advance of labelling rather than by an acquisition function.
Speciation was assigned geometrically, by proton ownership: a proton belongs to its nearest oxygen when that oxygen lies within a cutoff, and counting owners per oxygen identifies $\mathrm{H_3O^+}$ and $\mathrm{OH^-}$.
The first round drew neutral and ion-paired configurations from separate trajectories, since no single trajectory was found to hold both species in equilibrium.
The second round selected ion-paired configurations by the $\mathrm{H_3O^+}$--$\mathrm{OH^-}$ separation. \\

\noindent The ion pairs (ii) were successfully repaired by labelling.
The first round, $20$ configurations at $4.79$~V, showed the excess-electron error to be confined to ion-paired configurations and to grow with the separation of the $\mathrm{H_3O^+}$ and $\mathrm{OH^-}$ rather than saturating.
The second followed that trend directly, with $50$ configurations at $3.85$~V spanning separations from $4.2$ to $13.1$~\AA{}.
After retraining, the excess-electron error on held-out pair-separated configurations fell from $0.130$ to $0.055$~e, a factor of $2.4$, with no increase in error elsewhere in the dataset.

\clearpage

\section{Water structure using the mace-mh-1 foundation MLIP}
\label{app:mh1-water}

\begin{figure}[!htbp]
    \centering
    \includegraphics[width=\textwidth]{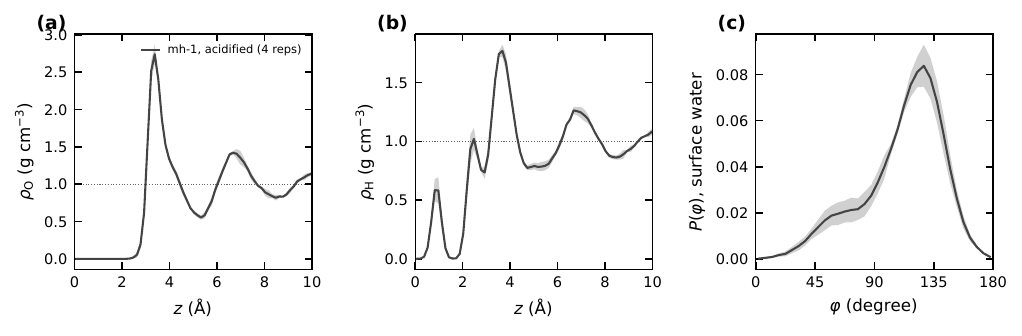}
    \caption{
    Interfacial water obtained using the foundation MLIP \texttt{mace-mh-1.model} (head \texttt{oc20\_usemppbe}) with D3(BJ) dispersion using Becke--Johnson damping.
    (a)~Oxygen and (b)~hydrogen mass-density profiles versus height $z$ above the outermost Au layer, and (c)~the orientation distribution of the water bisector for surface water ($z < 4$~\AA), defined as in the main text.
    }
    \label{fig:si-density-mh1}
\end{figure}
\noindent Figure~\ref{fig:si-density-mh1} shows the interfacial water structure obtained using the potential-\textit{independent} foundation MLIP \texttt{mace-mh-1.model} (head \texttt{oc20\_usemppbe}) on the $6\times4$ cell used for the production MD, and serves as a baseline for assessing the potential-\textit{dependent} changes.
The first oxygen peak is $2.74 \pm 0.13$~g\,cm$^{-3}$ at $z = 3.38$~\AA{} and the mean orientation is $\langle\varphi\rangle = 111^\circ$.
In-plane oxygen self-diffusion is $3.00 \pm 0.07 \times 10^{-5}$~cm$^2$/s, and the interior density ($8$--$16$~\AA{} above the outermost Au layer) is $0.970 \pm 0.006$~g\,cm$^{-3}$.
Uncertainties on these quantities are standard deviations across the four replicas.
Four trajectories were initialized using the same procedure as the production MD described in the Methods section of the main text.
Each trajectory was run for $542$~ps, with the first $200$~ps discarded before analysis.

\end{document}